\documentclass[a4paper]{aa}
\usepackage{isolatin1}      
\usepackage{graphicx}
\usepackage{natbib}
\begin{document}
   \title{The origin of Scorpius X-1}

   \author{I.F. Mirabel \inst{1,2}
	\and
		I. Rodrigues \inst{1}
		}

   \offprints{I. F. Mirabel, \email{fmirabel@cea.fr} }

   \institute{
	Service d'Astrophysique / CEA-Saclay, 91191 Gif-sur-Yvette, France
	\and
	Instituto de Astronomía y Física del Espacio/Conicet, Argentina
	       }

   \date{{\bf Published in Astronomy \& Astrophysics, 398, L25--L28 (2003)}}



   \abstract {We have used multi-wavelength 
observations of high precision to derive the space velocity and
compute the orbit around the Galactic Centre of the prototype X-ray
binary Scorpius X-1.  An origin in the local spiral arm of the Milky
Way is ruled out. The galactocentric kinematics of Scorpius X-1 is
similar to that of the most ancient stars and globular clusters of the
inner Galactic halo.  Most probably, this low-mass X-ray binary was
formed by a close encounter in a globular cluster.  However, it cannot
be ruled out that a natal supernova explosion launched Scorpius X-1
into an orbit like this from a birth place in the galactic bulge. In
any case, the Galactocentric orbit indicates that Scorpius X-1 was
formed more than 30 Myrs ago.
\keywords {stars: individual: Scorpius X-1 - X-rays: binaries -
Astrometry}
   }

   \maketitle
%

\section{Introduction}

Scorpius X-1 was the first discovered and it is the brightest
persistent extra-solar celestial X-ray source \citep{Giacconi}. It is
the prototype low-mass X-ray binary composed of a compact object and a
donor star that have masses of 1.4 M$_{\odot}$ and 0.42 M$_{\odot}$,
respectively \citep{Steeghs}. The orbital period is of 18.9 hr and the
compact object is accreting matter from the Roche lobe-filling
companion star. Although the mass of the compact object is consistent
with that of a neutron star, no thermonuclear explosion (X-ray burst
of type I) of the infalling matter on the putative hard surface of a
compact object has been reported so far.

The kinematical properties of Sco X-1 can help us in understanding its
origin. From the radial velocity, proper motion and distance of the
system, the space velocity can be derived. Thereafter, using a mass
model for the Galaxy, the galactic path can be tracked. This approach
was applied for first time to derive the Galactocentric orbits of the
halo black hole binary XTE J1118+480 \citep{Mirabelnat} and the
Galactic disk runaway black hole system GRO J1655-40 \citep{paperGRO}.

Scorpius X-1 is a microquasar \citep[for a review see][]{Mirabelrev},
namely, a source of relativistic jets with a compact radio core
associated to the accreting compact object \citep{Fomalont}. The most
accurate proper motion of X-ray binaries can be obtained using Very
Long Baseline Interferometry (VLBI) techniques at radio
wavelengths. In the case of Scorpius X-1, a precise distance of
2.8$\pm$0.3 kpc from the trigonometric parallax, and a proper motion
of the radio core have been determined using VLBI
\citep{Bradshaw}. Furthermore, an accurate value for the systemic
radial velocity was recently reported by \cite{Steeghs} from
spectroscopic observations at optical wavelengths.  Because of the
unprecedented precision in the distance, proper motion, and systemic
radial velocity, Scorpius X-1 is the X-ray binary for which we have
obtained the most accurate determination of the galactic path, among
the seven X-ray binary systems studied so far \citep{Mirabelcargese}.

\begin{figure*}[hhtb]
{\centering 
\resizebox*{1\textwidth}{!}{\rotatebox{00}{\includegraphics{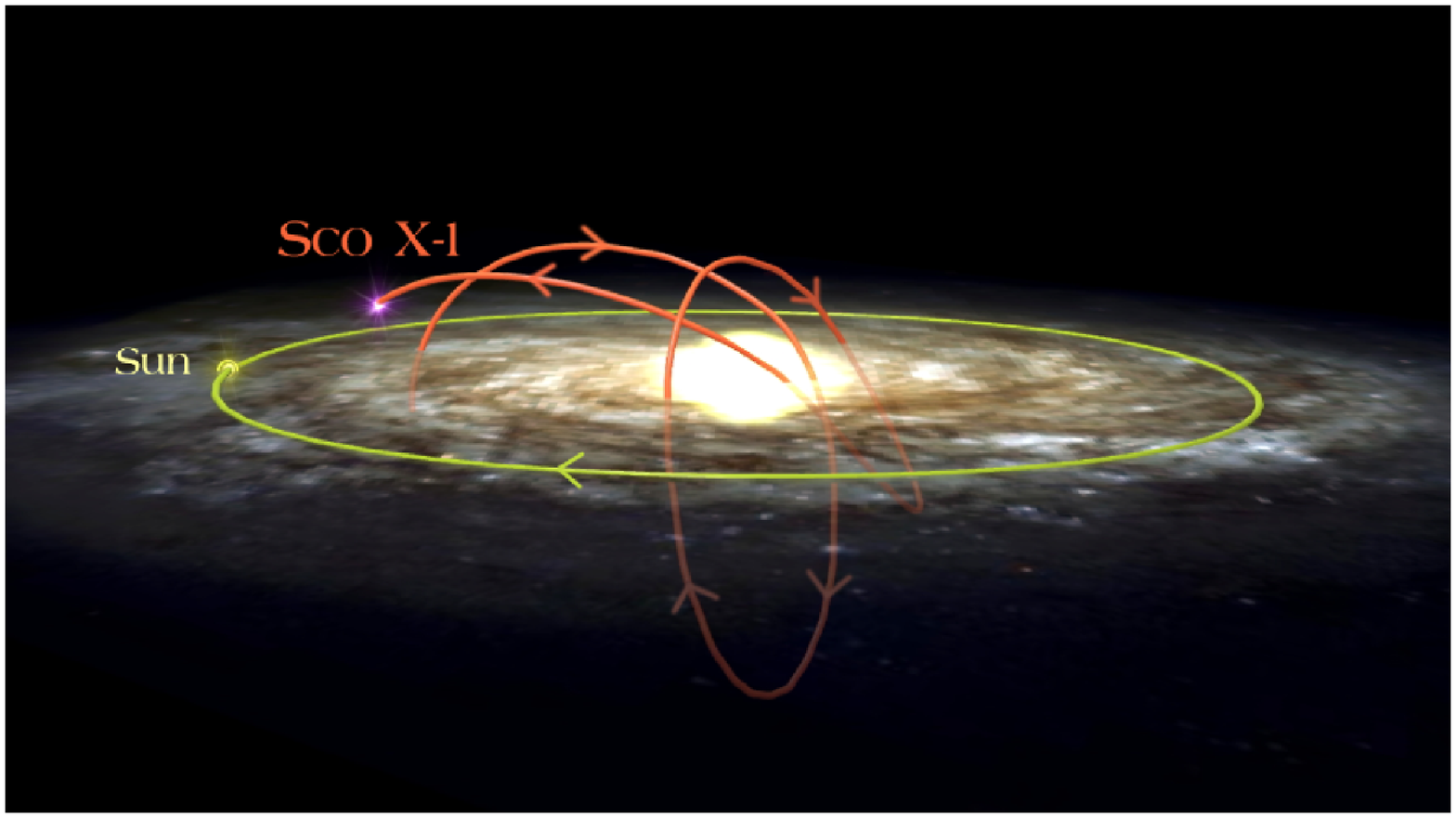}}}
\par}
\vskip 0.5 cm
{\centering 
\resizebox*{1\textwidth}{!}{\rotatebox{0}{\includegraphics{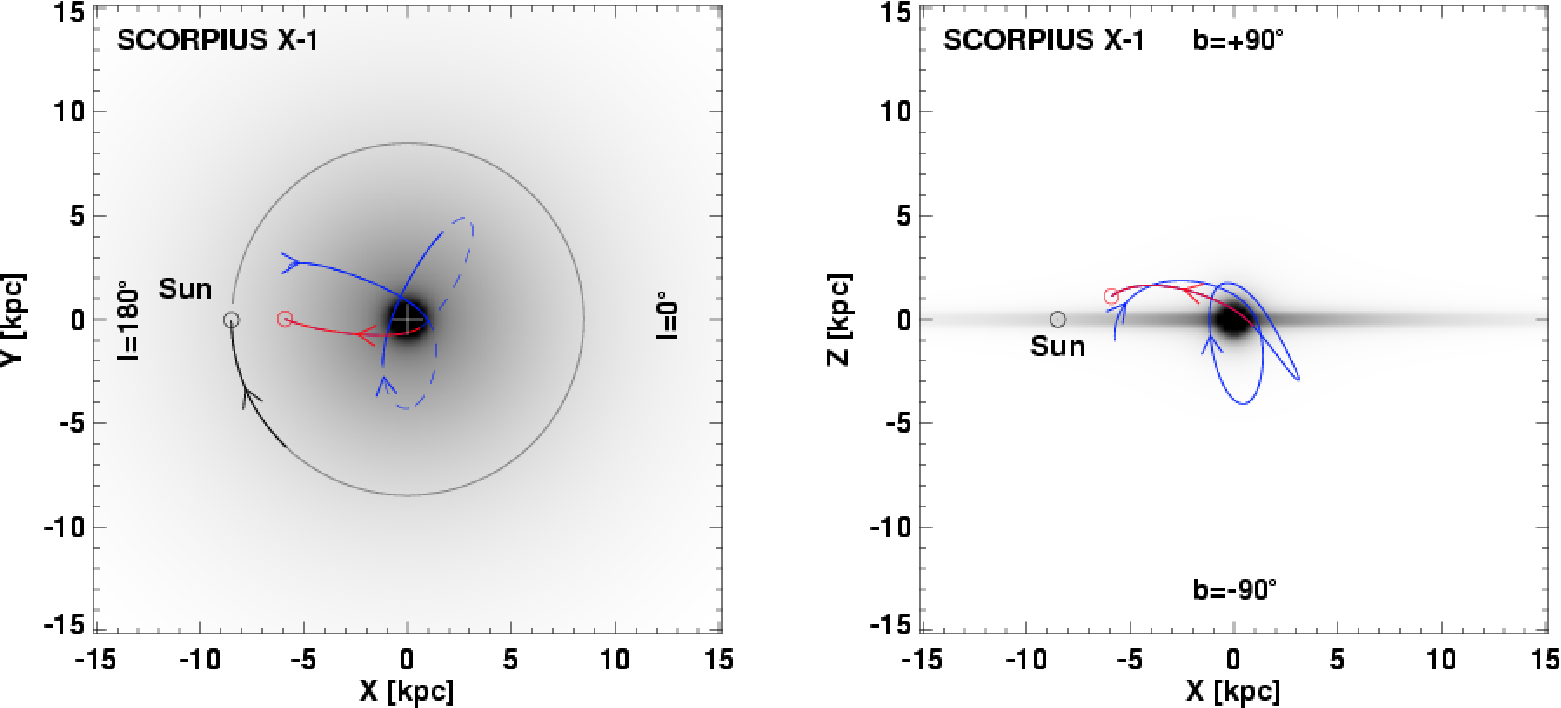}}}
\par}
\vskip 0.5 cm 
\caption{\textbf{Galactocentric orbit of Scorpius X-1}. 
In the top panel is traced in red color the orbit of Scorpius X-1
around the Galactic Centre and in yellow the orbit of the Sun for a
period of 230 Myrs. The lower two panels represent the view from above
the Galactic plane (left panel) and the side view (right panel). In
the left panel the continuous line represents the section of the orbit
in the Northern Galactic Hemisphere; the section of the orbit in the
Southern Hemisphere is shown by discontinuous trace. The orbit since
the last perigalactic distance of $\sim$500 pc that took place 30 Myrs
ago is shown in red color. An animation of the orbital motion of
Scorpius X-1 around the Galactic centre can be seen at:
http://www.iafe.uba.ar/astronomia/mirabel/mirabel.html
 \label{fig_orbit} }
\end{figure*}

\section{Space velocity and galactocentric orbit}

The position, distance, heliocentric radial velocity, and proper
motion are listed in Table~\ref{tab_data}. These data were used to
compute the galactic orbit of Scorpius X-1 using the galactic
gravitational potential model by \cite{stdpot}. The velocity
components U, V, and W, directed to the Galactic centre, rotation
direction, and North Galactic pole were derived using \cite{uvw}'s
equations of transformation, and assuming the sun moves with
components (U$_{\odot}$,V$_{\odot}$,W$_{\odot}$) = (9, 12, 7) km
s$^{-1}$ relative to the local standard of rest (LSR)
\citep{Mihalas}. The values of U, V, and W in Table~\ref{tab_data} are
rather different from the mean values that characterize the kinematics
of stars that belong to the thin and thick disk of the Galaxy
\citep{Chiba}.

\begin{table}[ht]
{\centering \begin{tabular}{|c|c|c|}
\hline 
{$l$ [$^{\circ }$]}&
{0.72}&
{}\\
\hline 
{$b$ [$^{\circ }$]}&
{+23.18}&
{}\\
\hline 
{D {[}kpc{]}}&
{2.8$\pm$0.3}&
{1}\\
\hline 
{V\( _\mathrm{helio} \) {[}km s\( ^{-1} \){]}}&
{-113.8$\pm$0.6}&
{2}\\
\hline 
{\( \mu  \)\( _{\alpha } \) {[}mas yr\( ^{-1} \){]}}&
{-6.88$\pm$0.07}&
{1}\\
\hline 
{\( \mu  \)\( _{\delta } \) {[}mas yr\( ^{-1} \){]}}&
{-12.02$\pm$0.16}&
{1}\\
\hline 
{U\( _\mathrm{LSR} \) {[}km s\( ^{-1} \){]}}&
{-93$\pm$2}&
{}\\
\hline 
{V\( _\mathrm{LSR} \) {[}km s\( ^{-1} \){]}}&
{-179$\pm$20}&
{}\\
\hline 
{W\( _\mathrm{LSR} \) {[}km s\( ^{-1} \){]}}&
{-74$\pm$4}&
{}\\
\hline 
{Galactocentric eccentricity}&
{0.87$\pm$0.05}&
{}\\
\hline 
{Z\( _\mathrm{max} \) [kpc]}&
{4.2$\pm$0.2}&
{}\\
\hline 
{\( D_\mathrm{peri} \) [kpc]}&
{0.5$\pm$0.1}&
{}\\
\hline 
{\( D_\mathrm{apo} \) [kpc]}&
{7.4$\pm$0.4}&
{}\\
\hline 
\end{tabular}\par}

\caption{\textbf{Data and computed galactocentric orbital parameters
for Scorpius\,X-1.} References given in $3^{rd}$ column are: 1 =
\cite{Bradshaw}, 2 = \cite{Steeghs} \label{tab_data}}

\end{table}

The parameters and graphic representations of the galactocentric orbit
are given at the end of Table~\ref{tab_data} and in
Figure~\ref{fig_orbit}, respectively. Because the distance, proper
motion and systemic radial velocity were determined with high
precision, the errors in the orbital parameters of the galactocentric
orbit of Scorpius X-1 are dominated by the uncertainties in the
Galactic potential model. The errors for the orbital parameters quoted
in Table~\ref{tab_data} are the deviations derived from comparing the
prescriptions by \cite{Flynn} and \cite{Marcio} with the one adopted
by us \citep{stdpot}. The computed orbits that result from the use of
the different current mass models for the Galaxy are very similar.

Scorpius X-1 does not participate in the Galactic rotation and its
kinematics is unlike those of stars that belong to the thin and thick
disk populations. Presently, Scorpius X-1 is moving towards the
Galactic plane with a free fall velocity of 74 km s$^{-1}$, and it is
receding from the Galactic Centre at a spatial speed of 125 km
s$^{-1}$.

\section{Discussion}

We first consider the possibility that Scorpius X-1 may be associated
with Gould's belt, which is a spur of the local spiral arm that has a
size of a few hundred parsecs. \cite{Bradshaw97} pointed out that 
Scorpius X-1 is beyond the
Sco-Cen association. The new refined distance of 2.8 kpc by
\cite{Bradshaw} locates the X-ray source farther beyond the edge of the
belt, which is expanding. Since Scorpius X-1 is approaching along the
line of sight with a velocity of 113.8 km s$^{-1}$ (see Table 1), the
possibility that Scorpius X-1 has been formed and ejected from Gould's
belt is ruled out.

Two possibilities remain: either Scorpius X-1 was kicked out into such
an orbit by the explosion of the massive progenitor of the compact
object, or it was formed in the halo itself. 30 Myr ago the binary was at 
a perigalactic distance of $\sim$500 pc with a velocity relative to the Galactic Centre of $\sim$480 km s$^{-1}$, implying -after subtraction 
of Galactic rotation- a minimum kick velocity of $\sim$230 km s$^{-1}$. 
This corresponds to a minimum linear momentum of $\sim$420 M$_{\odot}$ km s$^{-1}$. Although
large, this minimum linear momentum does not exceeds the maximum
linear momenta found among runaway solitary neutron stars and
millisecond pulsars \citep{Toscano}. Besides, \cite{Brandt} calculate  
that after a supernova explosion the resulting low mass X-ray binary can  
remain bound provided it does not attain a velocity larger than 
180$\pm$80 km s$^{-1}$. The minimum kick velocity for  
Scorpius X-1 is at the limit of the maximum velocity for which the binary can remain bound. Unfortunately, from the theoretical point of view, the timescales
after the explosion to achieve persistent Roche lobe overflow by the 
secondary and circularization of the
binary orbit, are both rather uncertain. Therefore, although unlikely, 
one cannot rule out the possibility that Scorpius X-1 was
launched from the Galactic bulge by the explosion of the progenitor of
the compact object.

The galactocentric orbit of Scorpius X-1 is consistent with an origin
in a globular cluster. The orbit is highly eccentric (e = 0.87), with
a perigalactic distance of 0.5 kpc, and a maximum height above the
plane of 4.2 kpc. It has comparable parameters to the orbits of
globular clusters of the inner halo, such as NGC 6121 and NGC 6712
\citep{Dauphole}. However,  we find that since the last perigalactic 
passage 30 Myrs ago, Scorpius X-1 did not intersect any known orbit
of a globular cluster.  

We point out that the majority of low-mass X-ray binaries of short
period are found in globular clusters. Because of the high stellar
densities and low velocity dispersions the cores of globular clusters
are the best sites in the Galaxy for the formation by close encounters 
of this class of
X-ray binaries. In this context, Scorpius X-1 most
probably was formed in the core of a globular cluster in a close
encounter of the compact object with a single star or with a binary
\citep{Verbunt}, being catapulted out of the core more than 30 Myrs ago.

\section{Conclusion}

Thanks to the unprecedented precision in the distance and proper
motion from VLBI measurements by \cite{Bradshaw}, and the accurate
systemic radial velocity measured with optical spectroscopy by
\cite{Steeghs}, a precise orbital motion of Scorpius X-1 around the
galactic centre was computed. The source reached a perigalactic
distance of 500 pc at a velocity of 480 km s$^{-1}$. Presently, it is
at a distance from the Sun of 2.8 kpc, falling to the Galactic plane
with a vertical speed of 74 km s$^{-1}$. A possible origin associated
to the Gould's belt can now be ruled out. Scorpius X-1 has an
eccentric orbit (e = 0.87) around the Galactic Centre, and the
kinematics is similar to that of the most ancient stars and globular
clusters of the inner halo. Most probably Scorpius X-1 has been formed
in the core of a globular cluster, as the majority of the low mass
X-ray binaries of short orbital period. However, we cannot rule out
that a natal supernova kick launched Scorpius X-1 into an orbit like
this from a birth place in the galactic bulge. In the context of these
two possibilities Scorpius X-1 must have been formed more than 30 Myrs
ago.

\begin{acknowledgements} We thank Jacques Paul and an anonymous
referee for very useful comments on the original manuscript. I.F.M. is
a member of the Consejo Nacional de Investigaciones Cient\'\i ficas y
T\'ecnicas of Argentina, and I.R. is a fellow of the Conselho Nacional
de Desenvolvimento Cien\'\i fico e Tecnol\'ogico of
Brazil. I.F.M. acknowledges support from PIP 0049/98 and Fundaci\'on
Antorchas.
\end{acknowledgements}

\bibliography{ScoX-1}

\bibliographystyle{aa}
  
\end{document}